\documentclass[twocolumn,showpacs,prl,aps,amsmath,amssymb]{revtex4}
\usepackage{graphicx}
\begin{document}

\title{Pressure induced magnetism in rotated graphene bilayers  }

\author{Felix Yndurain}
\email{felix.yndurain@uam.es.}
\affiliation{Departamento de F\'{i}sica de la Materia Condensada and 
Condensed Matter Physics Center (IFIMAC).
Universidad Aut\'{o}noma de Madrid. Cantoblanco. 28049 Madrid.
Spain.}
\date{\today}

\begin{abstract}
 
Using ab initio methods based on the density functional theory we show that rotated graphene bilayers at angles 
different  from the magic ones can have an electronic spectrum similar to those by applying moderate external pressures. We find that for an angle 
of $5.08^{\circ}$ and a pressure of 2.19 GPa the spin restricted spectrum displays a flat band at the Fermi level similar to the one found
at magic angles. In addition, the spin unrestricted calculations show a correlated ferromagnetic ground state with a total magnetic moment of 
3.7 $\mu_{B}$ per unit cell being mostly localized in the AA stacking region of the Moir\'{e} pattern. The possibility of antiferromagnetic order is considered but not calculated. Doping the system destroys the magnetic moments. The plausibility of BCS superconductivity in the doped system is analyzed.
\end{abstract}
 

\maketitle

The search of magnetism in graphene in its different forms has been the subject of intense effort in the last years. Based on Liebs's theorem \cite{Lieb} it is generally accepted that point defects induce localized magnetic moments in graphene \cite{Pereira,LopezSancho,Yazyev,Ugeda, Nair1, Yo}.
Most  attempts have been focused on single vacancies and atomic chemisorbed hydrogen. Magnetic moments induced by atomic hydrogen chemisorption in graphene has been well established \cite{Nosotros-Science} although the case of a single vacancy and other impurity is still an open question \cite{Yo-con-Juanjo, Pantelides, Ducastelle}.
However, very recently, the work of Cao et al. \cite{Jarillo-1} has open a new path to induce magnetism in defect free graphene systems. In their work  they find the presence of magnetic correlations in rotated graphene bilayers for the so called magic angles. Even more, Cao et al. \cite{Jarillo-2} find superconductivity with a  critical temperature of 1.7K in doped  
samples for a rotated angle of $1.1^{\circ}$. 

In this work, following the ideas of Cao et al. \cite{Jarillo-1, Jarillo-2}  we study the possibility of magnetic moments in rotated bilayers under pressure for small angles far for the magic ones. It has been shown recently \cite{Jarillo-Pressure} that small pressure 
can alleviate the stringent critical angle criteria. The efect of pressure on non twisted bilayers has also being studied recently \cite{Pressure-Balseiro}. In this work we show that even for angles as large as $5.08^{\circ}$ and applying pressures 
of the order of two GPa the eigenvalues spectrum displays a flat band at the Fermi level and consequently
magnetic correlations can take place.   

In order to study the geometrical and electronic structure of the different rotated  graphene bilayers we use a first principles density functional \cite {DFT1, DFT2} calculations using the SIESTA code \cite {Siesta1, Siesta2} which uses localized orbitals as basis functions \cite{Orbitals}. Most of the calculations are done using a double $\zeta$ basis set, non-local norm conserving pseudopotentials and, for the exchange correlation functional, we use the generalized gradient approximation (GGA)\cite{GGA} including van der Waals interaction with the functional developed by Berland and Hyldgaard \cite {BH} in the way implemented by Rom\'{a}n-P\'{e}rez and Soler \cite {Roman}. The results are checked with other functionals \cite{SM} finding that the results are essentially independent of the functional considered. The calculations are performed with stringent criteria in the electronic structure convergence; down to $10^{-5}$ in the density matrix, 2D Brillouin zone  sampling (up to 900 $k$-points), real space grid (energy cut-off of 400 Ryd) and equilibrium geometry (residual forces lower than $3\times10^{-2}$ eV/\AA). Due to the rapid variation of the density of states at the Fermi level, we used a polynomial smearing method \cite{smearing}.  

\begin{figure}[h]
\includegraphics[width=80mm]
{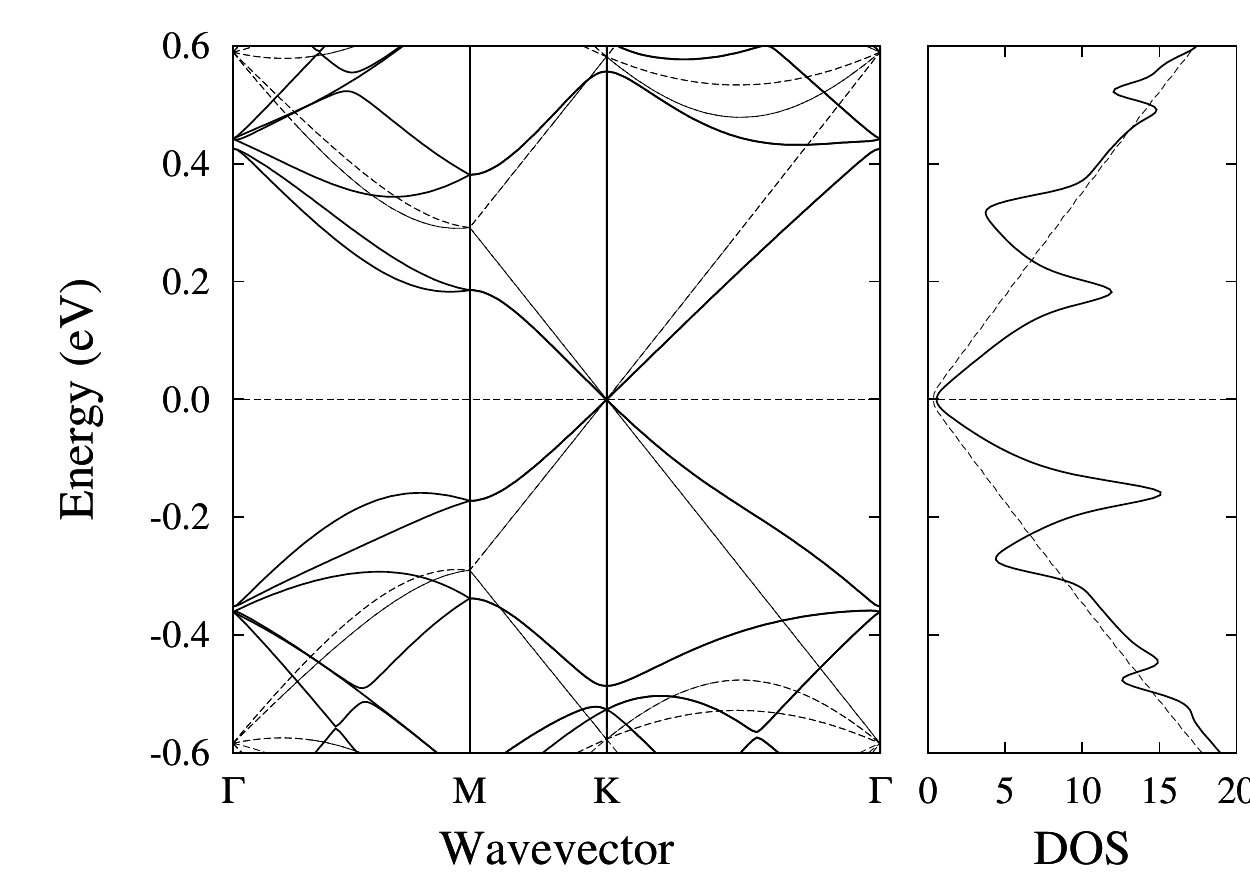}
\caption{Calculated band structure and density of states of a graphene bilayer rotated $3.89^{\circ}$ and a distance between the layers of 3.2 {\AA} (solid line). The broken line represents the same calculation when the two layers are decoupled. The model calculations are done using a single $\zeta$ basis.}
\label{Model-SZ}
\end{figure}

In Figure \ref{Model-SZ}  we show the results of a model using a  single $\zeta$ basis for two flat graphene layers rotated $3.89^{\circ}$ and at a 
distance of 3.2 {\AA}   as well as the results for the two layers being far apart. The main effect of the interaction between the layers is to open gaps of the degenerate bands at the point M of the Brillouin zone above and below the Dirac point as it is well known \cite{vanHove,Nosotros-bilayer}. Looking at this figure it is clear that there are two ways to merge the two empty and occupied van Hove logarithmic singularities in one single peak at the Dirac point; on one side by making the rotated angle smaller the saddle points get closer to the Dirac point and, at the magic angles, they merge in one single peak with the corresponding  flat band. On the other side, for a given angle, bringing the graphene layers closer (by applying pressure, for instance) the interaction between the layers is larger (the so called $t_{\theta}$ parameter \cite{Laissardiere-PRB}) opening larger gaps and therefore bringing the saddle points to the Dirac point.

We have then calculated the electronic structure and geometrical equilibrium configuration for two graphene layers rotated $5.08^{\circ}$.
In this case we use an optimized double $\zeta$ basis and the van der Waals functional of Berland and Hyldgaard \cite {BH}. The calculation entails a unit cell of 508 atoms in the limit of the capacities of the SIESTA code in its present form. 
This angle of $5.08^{\circ}$ corresponds to a 6$\_$7 unit cell in the notation of Laissardiere et al. \cite{Laissardiere1}; namely
callying $\vec{a_{1}}$ and  $\vec{a_{2}}$ the primitive lattice vectors, the new ones for the (n$\_$m) unit cell are: $\vec{t}=n \vec{a_{1}}+m\vec{a_{2}}$ and $\vec{t'}=-m \vec{a_{1}}+(n+m)\vec{a_{2}}$ .  The rotation angle is then given by
$cos\theta =\frac{n^{2}+4nm+m^{2}}{2(n^{2}+nm+m^{2})}$.

\begin{figure}[h]
\includegraphics[width=80mm]
{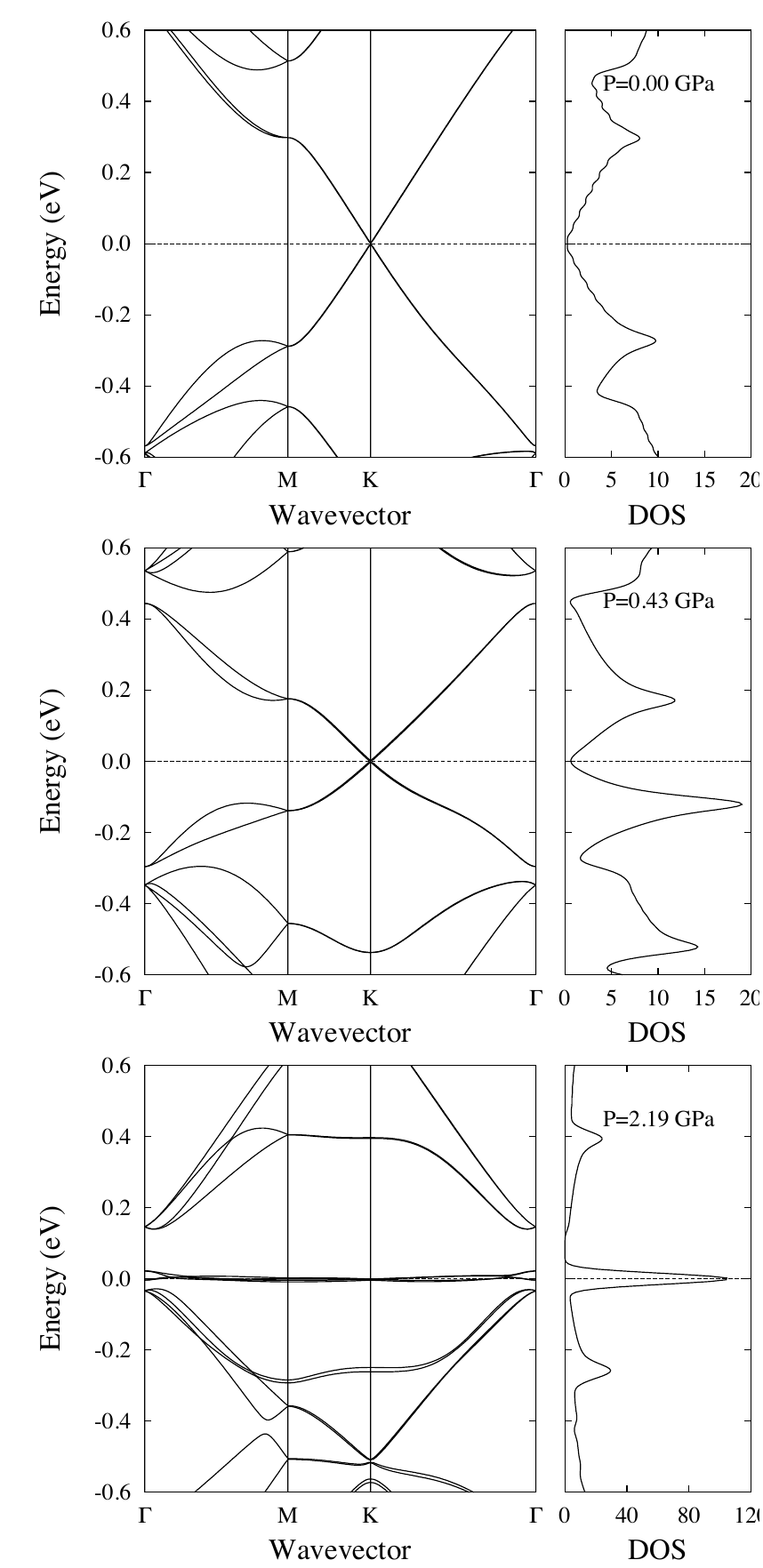}
\caption{Spin restricted calculated of the band structure and densities of states of a 6$\_$7 unit cell corresponding to an $5.08^{\circ}$ angle between the two layers. The different panels show the results for different pressures. A small (0.02 eV) gaussian broadening is included in the density of states calculation for presentation purposes. }
\label{Bands-vs-Pressure}
\end{figure}

The calculations were performed for various different separation between the layers. In Figure \ref{Bands-vs-Pressure} we show the results for 0.0, 0.4 and 0.8 {\AA} compression with respect to the equilibrium geometry which correspond to external pressures of 0.0, 0.43 and 2.19 GPa respectively. In this case the calculations are spin-restricted with the constraint of equal population of spin up and spin down population. The calculations are such that, for the zero pressure case, the full geometry is relaxed to residual forces lower than $2\times10^{-2}$ eV/\AA. For non zero pressures the layers are rigidity shifted without any further relaxation. The corresponding pressure is calculated afterwards. The shown results are essentially independent of the functional used as well as the relaxation  within the layers \cite{SM}.

The results of Figure \ref{Bands-vs-Pressure} are remarkably similar to those of magical angles. The advantage in this case, in terms of computational effort, is that a much smaller unit cell is required and, in addition, the external parameter pressure is more easily tuned than the angle between the layers.

The results of this figure shows how decreasing the distance between the layers and therefore increasing the external pressure increases the 
gap that gives rise to the van Hove logarithmic singularities.  These singularities become sharper and eventually collapse in in one degenerated band and a very sharp peak
in the density of states. The peak is similar to the one obtained by bringing the angle to a magic one. It would be interesting to study how the Fermi surface varies with the pressure, this is beyond the scope of the present work but it will be considered in the future.

\begin{figure}[h]
\includegraphics[width=80mm]
{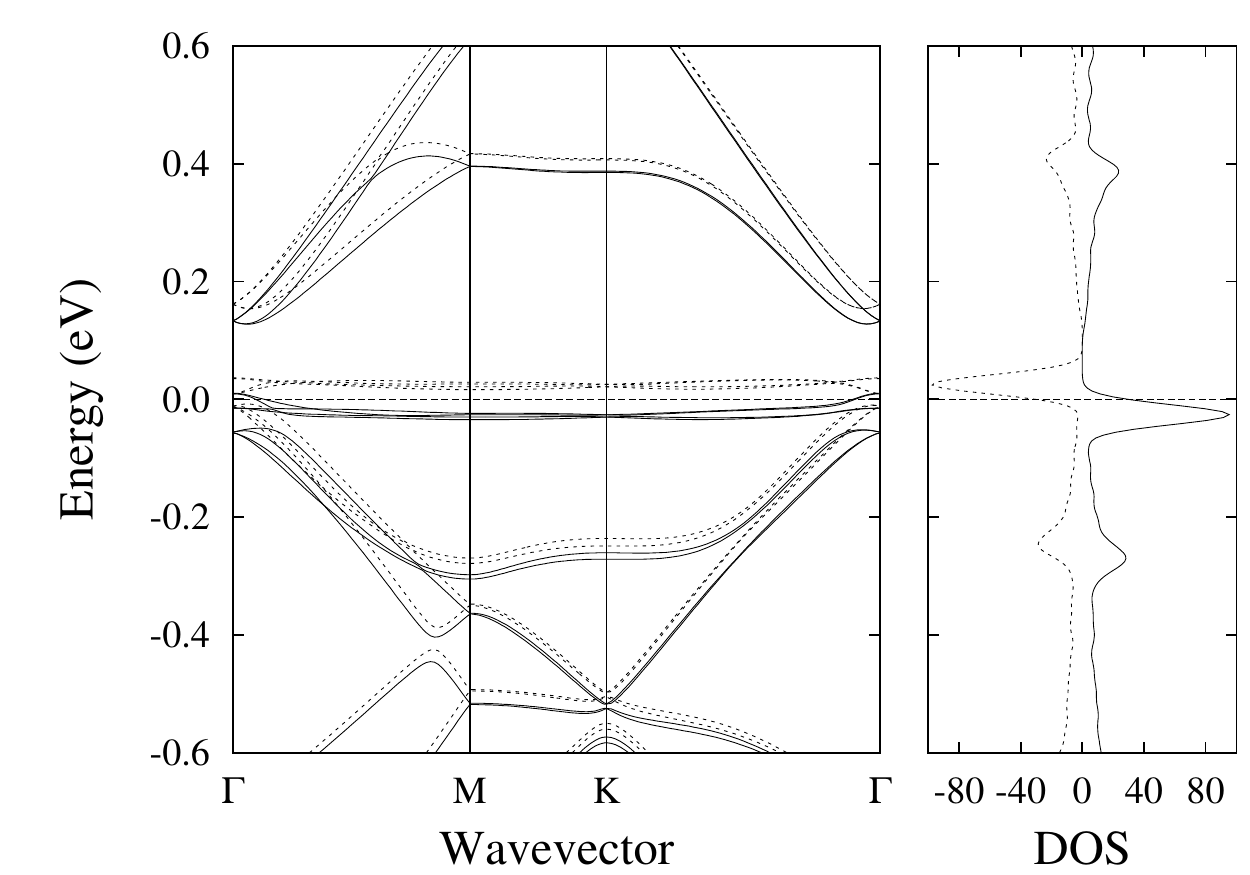}
\caption{Spin-resolved band structure (left panel) and density of states (right panel) of a 6$\_$7 rotated bilayer ($5.08^{\circ}$) under a 2.19 GPa pressure. The horizontal dotted line indicates the Fermi energy. A small (0.02 eV) gaussian broadening is included in the density of states for presentation purposes. Spin up (down) states are presented by heavy (broken) lines.}
\label{Bands-6_7-Mag}
\end{figure}

At this point it is interesting to study the stability of the system with respect to broken symmetries in particular to the spin one. We have therefore done a spin unrestricted calculation for the pressure of 2.19 Gpa. For other pressures shown we indeed do not expect any magnetic instability.
The results of the band structure and density of states are shown in Figure \ref{Bands-6_7-Mag}. We immediately see the spin up spin-down bands splitting. The total magnetic moment is 3.7  $\mu_{B}$ close to the nominal value of 4 $\mu_{B}$ corresponding to degenerate 
flat bands. 

\begin{figure}[h]
\includegraphics[width=90mm]
{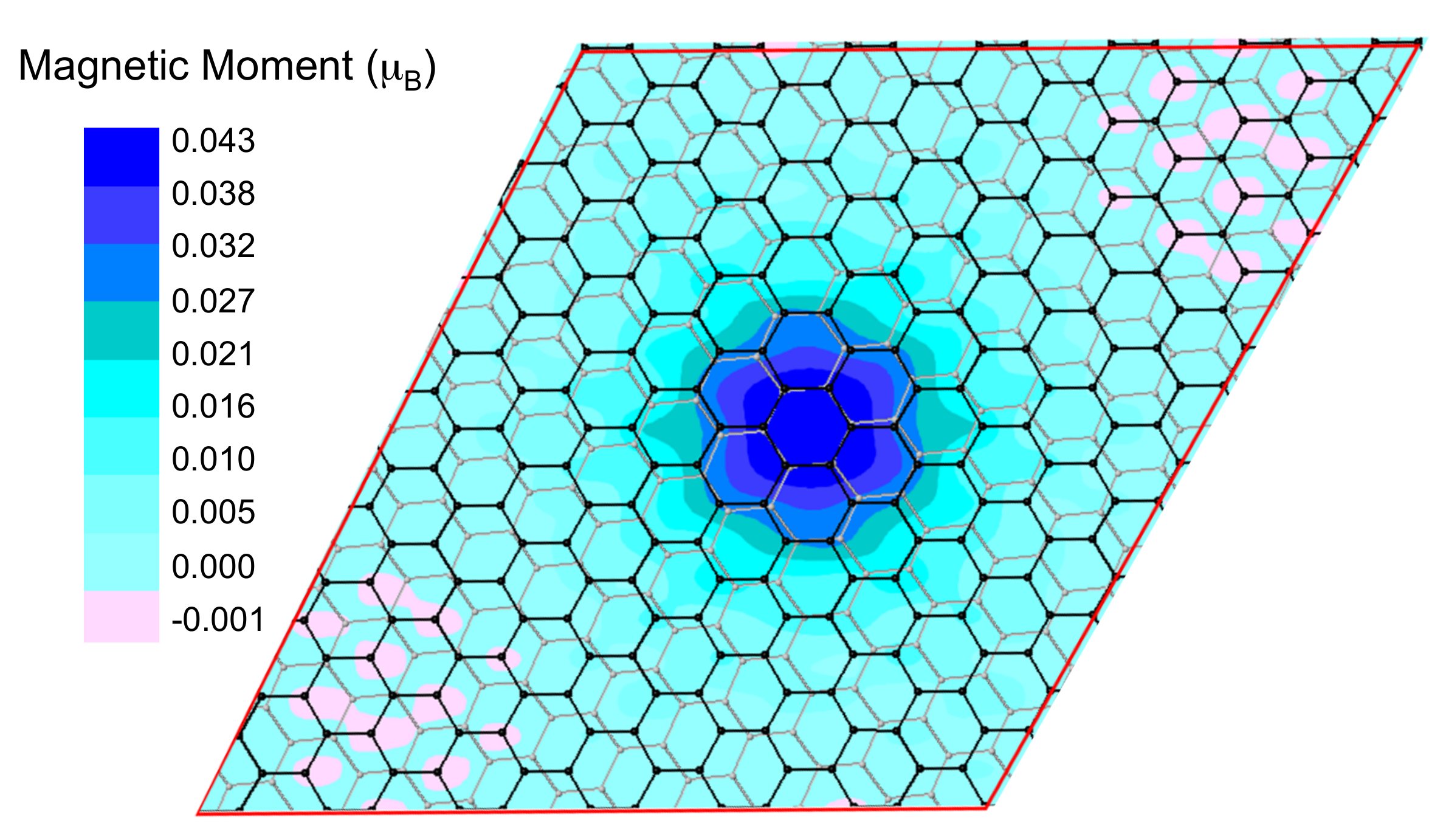}
\caption{(color on line) Magnetic moments map of the top layer of a 6$\_$7 bilayer, layers rotated $5.08^{\circ}$,  under a pressure of 2.19 GPa. One primitive unit cell is shown.The intensity of the color indicates size of the magnetic moment in the atoms. White areas indicate negative values.}
\label{Magnetic-Moment-Map}
\end{figure}

It is interesting to study how these 3.7 $\mu_{B}$ are distributed in the Moir\'{e} pattern.
In Figure \ref{Magnetic-Moment-Map} we show, in units of $\mu_{B}$, the magnetic moment at one layer of the compressed 
6$\_$7 bilayer (for the other layer the results are indeed similar). We observe that the magnetic moment appears mainly in the AA stacking region, whereas in the rest
of the layer the magnetic moment is very small. This is not surprising since the states associated to the flat bands 
are mainly localized at this region\cite{SM} as discussed in detail by  Laissardiere et al. \cite{Laissardiere1} using a tight binding Hamiltonian and by 
Lopes dos Santos et al. \cite{Continuum-1, Continuum-2} within the continuum model. It is worth to comment that in the AB stacking zone the magnetic configuration is antiferromagnetic. This could be an indication that the magnetic configuration of second neighbor
AA zones could be antiferromagnetic maintaining the size of the moments in each AA stacking region. We cannot check this point
for the time being since it would require  a four times larger unit cell of 2032 atoms which exceeds the capacity of our code.

\begin{figure}[h]
\includegraphics[width=105mm]
{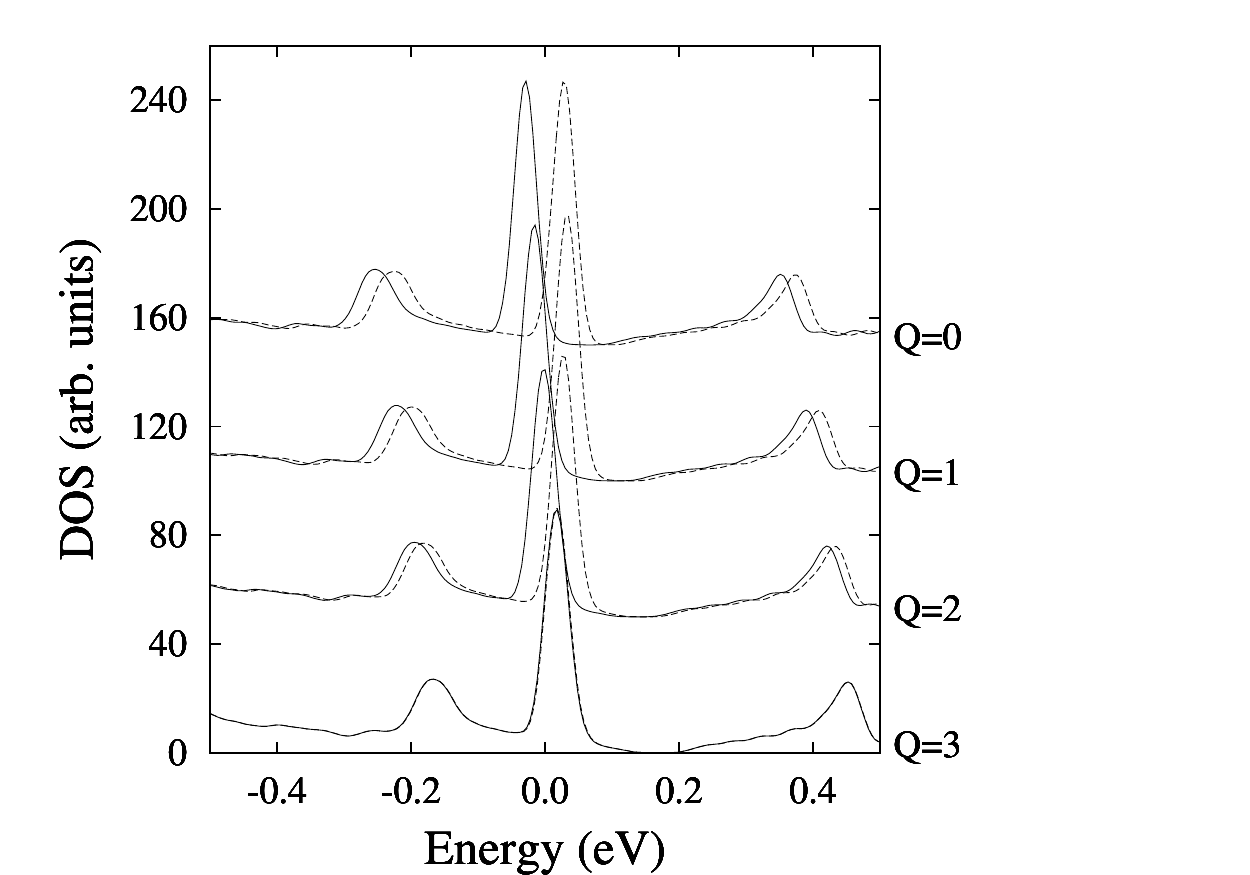}
\caption{Density of states versus energy relative to $E_{F}$ of a 6$\_$7 bilayer,$5.08^{\circ}$,  under a pressure of 2.19 GPa. for different hole doping.
Solid and broken lines represent spin up and spin down states respectively. A small (0.02 eV) gaussian broadening is included for presentation purposes.}
\label{DOS-vs-Doping}
\end{figure}

The recent experiments of Cao et al. \cite{Jarillo-1, Jarillo-2} indicate a very rich phase diagram when dealing with doped samples. We have therefore studied how  magnetism is affected by the presence of extra electrons or holes. 
In Figure \ref{DOS-vs-Doping} we show how magnetism disappears upon hole doping. We show the spin resolved densities of states for 0 to 3 extra holes. Magnetism completely vanishes when adding three holes while displaying a very high density of states at the Fermi level due to the narrowness of the peak. At this point we can speculate with the possibility of BCS superconductivity in doped samples with a high density of states at the Fermi level, namely at the paramagnetic flat band (like for the case Q=3 in  Figure \ref{DOS-vs-Doping} ). We, following the works of An and Pickett \cite{Pickett} and Boeri et al. \cite{Andersen} for MgB$_2$ and doped diamond, have estimated the deformation potential  associated to the flat band in the presence of a inter-layer breathing phonon mode. We have calculated the phonon frequency of this mode for the 0 pressure case to be 110 cm$^{-1}$ close to the experimental value \cite{Phonons} and to the value of the AB structure. One rough way to estimate the electron phonon interaction $\lambda$ by  
$\lambda=N(E_{F})\frac{D^{2}}{M\omega ^{2}}$
where D is the deformation potential. We find a density of state at the Fermi level of 
N=0.47 states/eV/f.u. The deformation potential can be estimated by means of the variation of
the energy position of the van Hove logarithmic
peaks with the amplitude of the mode. $D=\frac{\Delta E}{\Delta u}$ where E is the energy of the 
logarithmic peak and u is the phonon amplitude. We find for the deformation potential a value of the order of 0.6 eV/{\AA}. Putting all these numbers together we end up with a value for the electron-phonon interaction parameter 
$\lambda$=0.057. This value is compatible with a non-zero transition temperature. The result suggests that it is worth further investigation in this direction.

In summary, we have studied the appearance of magnetic moments in compressed rotated graphene bilayers. From our calculations we can conclude the following: 

\textit{i)} For a given rotation angle between the layers, bringing the graphene layers closer with respect to the equilibrium distance, shifts the logarithmic van Hove singularities towards the Dirac point eventually merging in one single peak. 

\textit{ii)} For a given distance between the layers equivalent to a  2.19 GPa pressure for layers rotated $5.08^{\circ}$, the system becomes magnetic with a magnetic moment close to 4 $\mu_{B}$ per unit cell in the Moir\'{e}. 

 \textit{iii)} The magnetic moments are mainly localized at the AA stacking zones. This could be anticipated since states associated to the flat bands are localized in these zones. 
 
 \textit{iv)} The possibility of antiferromagnetic order as well as conventional superconductivity in doped samples is only suggested in our calculations and deserves further research.
     
 \textit{v)} No evidence of either charge or spin density waves is found.
 

\subsection{Acknowledgements} I  would like to thank Profs. I. Brihuega, J. M. G\'{o}mez-Rodr\'{i}guez,  G. G\'{o}mez-Santos  and J. M. Soler for many lively and illuminating discussions. I am indebted to  Spanish Ministry of Science and Innovation for financial support through grant FIS2015-64886-C5-5-P.
\bibliographystyle{apsrev}

\begin{thebibliography}{34}
\expandafter\ifx\csname natexlab\endcsname\relax\def\natexlab#1{#1}\fi
\expandafter\ifx\csname bibnamefont\endcsname\relax
  \def\bibnamefont#1{#1}\fi
\expandafter\ifx\csname bibfnamefont\endcsname\relax
  \def\bibfnamefont#1{#1}\fi
\expandafter\ifx\csname citenamefont\endcsname\relax
  \def\citenamefont#1{#1}\fi
\expandafter\ifx\csname url\endcsname\relax
  \def\url#1{\texttt{#1}}\fi
\expandafter\ifx\csname urlprefix\endcsname\relax\def\urlprefix{URL }\fi
\providecommand{\bibinfo}[2]{#2}
\providecommand{\eprint}[2][]{\url{#2}}

\bibitem[{\citenamefont{Lieb}(1989)}]{Lieb}
\bibinfo{author}{\bibfnamefont{E.~H.} \bibnamefont{Lieb}},
  \bibinfo{journal}{Phys. Rev. Lett.} \textbf{\bibinfo{volume}{62}},
  \bibinfo{pages}{1201} (\bibinfo{year}{1989}).

\bibitem[{\citenamefont{Pereira et~al.}(2006)\citenamefont{Pereira, Guinea, dos
  Santos, Peres, and Neto}}]{Pereira}
\bibinfo{author}{\bibfnamefont{V.~M.} \bibnamefont{Pereira}},
  \bibinfo{author}{\bibfnamefont{F.}~\bibnamefont{Guinea}},
  \bibinfo{author}{\bibfnamefont{J.~M. B.} \bibnamefont{Lopes dos Santos}},
  \bibinfo{author}{\bibfnamefont{N.~M.~R.} \bibnamefont{Peres}},
  \bibnamefont{and} \bibinfo{author}{\bibfnamefont{A.~H.}
  \bibnamefont{Castro~Neto}}, \bibinfo{journal}{Phys. Rev. Lett.}
  \textbf{\bibinfo{volume}{96}}, \bibinfo{pages}{036801}
  (\bibinfo{year}{2006}).

\bibitem[{\citenamefont{L\'{o}pez-Sancho
  et~al.}(209)\citenamefont{L\'{o}pez-Sancho, de~Juan, and
  Vozmediano}}]{LopezSancho}
\bibinfo{author}{\bibfnamefont{M.~P.} \bibnamefont{L\'{o}pez-Sancho}},
  \bibinfo{author}{\bibfnamefont{F.}~\bibnamefont{de~Juan}}, \bibnamefont{and}
  \bibinfo{author}{\bibfnamefont{M.~A.~H.} \bibnamefont{Vozmediano}},
  \bibinfo{journal}{Phys. Rev. B} \textbf{\bibinfo{volume}{79}},
  \bibinfo{pages}{075413} (\bibinfo{year}{209}).

\bibitem[{\citenamefont{Yazyev and Helm}(2007)}]{Yazyev}
\bibinfo{author}{\bibfnamefont{O.~V.} \bibnamefont{Yazyev}} \bibnamefont{and}
  \bibinfo{author}{\bibfnamefont{L.}~\bibnamefont{Helm}},
  \bibinfo{journal}{Phys. Rev. B} \textbf{\bibinfo{volume}{75}},
  \bibinfo{pages}{125408} (\bibinfo{year}{2007}).

\bibitem[{\citenamefont{Ugeda et~al.}(2010)\citenamefont{Ugeda, Brihuega,
  Guinea, and G\'{o}mez-Rodr\'{i}guez}}]{Ugeda}
\bibinfo{author}{\bibfnamefont{M.~M.} \bibnamefont{Ugeda}},
  \bibinfo{author}{\bibfnamefont{I.}~\bibnamefont{Brihuega}},
  \bibinfo{author}{\bibfnamefont{F.}~\bibnamefont{Guinea}}, \bibnamefont{and}
  \bibinfo{author}{\bibfnamefont{J.~M.} \bibnamefont{G\'{o}mez-Rodr\'{i}guez}},
  \bibinfo{journal}{Phys. Rev. Lett.} \textbf{\bibinfo{volume}{104}},
  \bibinfo{pages}{096804} (\bibinfo{year}{2010}).

\bibitem[{\citenamefont{Nair et~al.}(2013)\citenamefont{Nair, Tsai, Sepioni,
  Lehtinen, Keinonen, Krasheninnikov, Neto, Katsnelson, and Geim}}]{Nair1}
\bibinfo{author}{\bibfnamefont{R.~R.} \bibnamefont{Nair}},
  \bibinfo{author}{\bibfnamefont{I.~L.} \bibnamefont{Tsai}},
  \bibinfo{author}{\bibfnamefont{M.}~\bibnamefont{Sepioni}},
  \bibinfo{author}{\bibfnamefont{O.}~\bibnamefont{Lehtinen}},
  \bibinfo{author}{\bibfnamefont{J.}~\bibnamefont{Keinonen}},
  \bibinfo{author}{\bibfnamefont{A.~V.} \bibnamefont{Krasheninnikov}},
  \bibinfo{author}{\bibfnamefont{A.~H.~C.} \bibnamefont{Neto}},
  \bibinfo{author}{\bibfnamefont{M.~I.} \bibnamefont{Katsnelson}},
  \bibnamefont{and} \bibinfo{author}{\bibfnamefont{A.~K.} \bibnamefont{Geim}},
  \bibinfo{journal}{Nature Communications} \textbf{\bibinfo{volume}{4}},
  \bibinfo{pages}{2010} (\bibinfo{year}{2013}).

\bibitem[{\citenamefont{Yndurain}(2014)}]{Yo}
\bibinfo{author}{\bibfnamefont{F.}~\bibnamefont{Yndurain}},
  \bibinfo{journal}{Phys. Rev. B} \textbf{\bibinfo{volume}{90}},
  \bibinfo{pages}{245420} (\bibinfo{year}{2014}).

\bibitem[{\citenamefont{Gonz\'{a}lez-Herrero
  et~al.}(2016)\citenamefont{Gonz\'{a}lez-Herrero, G\'{o}mez-Rodr\'{i}guez,
  Mallet, Moaied, Palacios, Salgado, Ugeda, Veullien, Yndurain, and
  Brihuega}}]{Nosotros-Science}
\bibinfo{author}{\bibfnamefont{H.}~\bibnamefont{Gonz\'{a}lez-Herrero}},
  \bibinfo{author}{\bibfnamefont{J.}~\bibnamefont{G\'{o}mez-Rodr\'{i}guez}},
  \bibinfo{author}{\bibfnamefont{P.}~\bibnamefont{Mallet}},
  \bibinfo{author}{\bibfnamefont{M.}~\bibnamefont{Moaied}},
  \bibinfo{author}{\bibfnamefont{J.}~\bibnamefont{Palacios}},
  \bibinfo{author}{\bibfnamefont{C.}~\bibnamefont{Salgado}},
  \bibinfo{author}{\bibfnamefont{M.}~\bibnamefont{Ugeda}},
  \bibinfo{author}{\bibfnamefont{J.-Y.} \bibnamefont{Veullien}},
  \bibinfo{author}{\bibfnamefont{F.}~\bibnamefont{Yndurain}}, \bibnamefont{and}
  \bibinfo{author}{\bibfnamefont{I.}~\bibnamefont{Brihuega}},
  \bibinfo{journal}{Science} \textbf{\bibinfo{volume}{352}},
  \bibinfo{pages}{437} (\bibinfo{year}{2016}).

\bibitem[{\citenamefont{Palacios and Yndurain}(2012)}]{Yo-con-Juanjo}
\bibinfo{author}{\bibfnamefont{J.~J.} \bibnamefont{Palacios}} \bibnamefont{and}
  \bibinfo{author}{\bibfnamefont{F.}~\bibnamefont{Yndurain}},
  \bibinfo{journal}{Phys. Rev. B} \textbf{\bibinfo{volume}{85}},
  \bibinfo{pages}{245443} (\bibinfo{year}{2012}).

\bibitem[{\citenamefont{Wang and Pantelides}(2012)}]{Pantelides}
\bibinfo{author}{\bibfnamefont{B.}~\bibnamefont{Wang}} \bibnamefont{and}
  \bibinfo{author}{\bibfnamefont{S.~T.} \bibnamefont{Pantelides}},
  \bibinfo{journal}{Phys. Rev. B} \textbf{\bibinfo{volume}{86}},
  \bibinfo{pages}{165438} (\bibinfo{year}{2012}).

\bibitem[{\citenamefont{Ducastelle}(2013)}]{Ducastelle}
\bibinfo{author}{\bibfnamefont{F.}~\bibnamefont{Ducastelle}},
  \bibinfo{journal}{Phys. Rev. B} \textbf{\bibinfo{volume}{88}},
  \bibinfo{pages}{075413} (\bibinfo{year}{2013}).

\bibitem[{\citenamefont{Cao et~al.}(2018{\natexlab{a}})\citenamefont{Cao,
  Fatemi, Fang, Watanabe, Taniguchi, Kaxiras, and Jarillo-Herrero}}]{Jarillo-1}
\bibinfo{author}{\bibfnamefont{Y.}~\bibnamefont{Cao}},
  \bibinfo{author}{\bibfnamefont{V.}~\bibnamefont{Fatemi}},
  \bibinfo{author}{\bibfnamefont{S.}~\bibnamefont{Fang}},
  \bibinfo{author}{\bibfnamefont{K.}~\bibnamefont{Watanabe}},
  \bibinfo{author}{\bibfnamefont{T.}~\bibnamefont{Taniguchi}},
  \bibinfo{author}{\bibfnamefont{E.}~\bibnamefont{Kaxiras}}, \bibnamefont{and}
  \bibinfo{author}{\bibfnamefont{P.}~\bibnamefont{Jarillo-Herrero}},
  \bibinfo{journal}{Nature (London)} \textbf{\bibinfo{volume}{556}},
  \bibinfo{pages}{43} (\bibinfo{year}{2018}{\natexlab{a}}).

\bibitem[{\citenamefont{Cao et~al.}(2018{\natexlab{b}})\citenamefont{Cao,
  Fatemi, Demi, Fang, Tomarken, Luo, Sanchez-Yamagishi, Watanabe, Taniguchi,
  Kaxiras et~al.}}]{Jarillo-2}
\bibinfo{author}{\bibfnamefont{Y.}~\bibnamefont{Cao}},
  \bibinfo{author}{\bibfnamefont{V.}~\bibnamefont{Fatemi}},
  \bibinfo{author}{\bibfnamefont{A.}~\bibnamefont{Demi}},
  \bibinfo{author}{\bibfnamefont{S.}~\bibnamefont{Fang}},
  \bibinfo{author}{\bibfnamefont{S.}~\bibnamefont{Tomarken}},
  \bibinfo{author}{\bibfnamefont{J.~Y.} \bibnamefont{Luo}},
  \bibinfo{author}{\bibfnamefont{J.}~\bibnamefont{Sanchez-Yamagishi}},
  \bibinfo{author}{\bibfnamefont{K.}~\bibnamefont{Watanabe}},
  \bibinfo{author}{\bibfnamefont{T.}~\bibnamefont{Taniguchi}},
  \bibinfo{author}{\bibfnamefont{E.}~\bibnamefont{Kaxiras}},
  \bibnamefont{et~al.}, \bibinfo{journal}{Nature (London)}
  \textbf{\bibinfo{volume}{556}}, \bibinfo{pages}{80}
  (\bibinfo{year}{2018}{\natexlab{b}}).

\bibitem[{\citenamefont{Carr et~al.}(2018)\citenamefont{Carr, Fang,
  Jarillo-Herrero, and Kaxiras}}]{Jarillo-Pressure}
\bibinfo{author}{\bibfnamefont{S.}~\bibnamefont{Carr}},
  \bibinfo{author}{\bibfnamefont{S.}~\bibnamefont{Fang}},
  \bibinfo{author}{\bibfnamefont{P.}~\bibnamefont{Jarillo-Herrero}},
  \bibnamefont{and} \bibinfo{author}{\bibfnamefont{E.}~\bibnamefont{Kaxiras}},
  \bibinfo{journal}{Phys. Rev. B} \textbf{\bibinfo{volume}{98}},
  \bibinfo{pages}{085144} (\bibinfo{year}{2018}).

\bibitem[{\citenamefont{Munoz et~al.}(2016)\citenamefont{Munoz, Collado, Usaj,
  Sofo, and Balseiro}}]{Pressure-Balseiro}
\bibinfo{author}{\bibfnamefont{F.}~\bibnamefont{Munoz}},
  \bibinfo{author}{\bibfnamefont{H.~O.} \bibnamefont{Collado}},
  \bibinfo{author}{\bibfnamefont{G.}~\bibnamefont{Usaj}},
  \bibinfo{author}{\bibfnamefont{J.}~\bibnamefont{Sofo}}, \bibnamefont{and}
  \bibinfo{author}{\bibfnamefont{C.}~\bibnamefont{Balseiro}},
  \bibinfo{journal}{Phys. Rev. B} \textbf{\bibinfo{volume}{93}},
  \bibinfo{pages}{235443} (\bibinfo{year}{2016}).

\bibitem[{\citenamefont{Hohenberg and Kohn}(1964)}]{DFT1}
\bibinfo{author}{\bibfnamefont{P.}~\bibnamefont{Hohenberg}} \bibnamefont{and}
  \bibinfo{author}{\bibfnamefont{W.}~\bibnamefont{Kohn}},
  \bibinfo{journal}{Phys. Rev.} \textbf{\bibinfo{volume}{136}},
  \bibinfo{pages}{B864} (\bibinfo{year}{1964}).

\bibitem[{\citenamefont{Kohn and Sham}(1965)}]{DFT2}
\bibinfo{author}{\bibfnamefont{W.}~\bibnamefont{Kohn}} \bibnamefont{and}
  \bibinfo{author}{\bibfnamefont{L.~J.} \bibnamefont{Sham}},
  \bibinfo{journal}{Phys. Rev.} \textbf{\bibinfo{volume}{140}},
  \bibinfo{pages}{A1133} (\bibinfo{year}{1965}).

\bibitem[{\citenamefont{Ordej\'{o}n et~al.}(1996)\citenamefont{Ordej\'{o}n,
  Artacho, and Soler}}]{Siesta1}
\bibinfo{author}{\bibfnamefont{P.}~\bibnamefont{Ordej\'{o}n}},
  \bibinfo{author}{\bibfnamefont{E.}~\bibnamefont{Artacho}}, \bibnamefont{and}
  \bibinfo{author}{\bibfnamefont{J.~M.} \bibnamefont{Soler}},
  \bibinfo{journal}{Phys. Rev. B} \textbf{\bibinfo{volume}{53}},
  \bibinfo{pages}{R10441} (\bibinfo{year}{1996}).

\bibitem[{\citenamefont{Soler et~al.}(2002)\citenamefont{Soler, Artacho, Gale,
  Garc\'{i}a, Junquera, Ordej\'{o}n, and Sanchez-Portal}}]{Siesta2}
\bibinfo{author}{\bibfnamefont{J.}~\bibnamefont{Soler}},
  \bibinfo{author}{\bibfnamefont{E.}~\bibnamefont{Artacho}},
  \bibinfo{author}{\bibfnamefont{J.}~\bibnamefont{Gale}},
  \bibinfo{author}{\bibfnamefont{A.}~\bibnamefont{Garc\'{i}a}},
  \bibinfo{author}{\bibfnamefont{J.}~\bibnamefont{Junquera}},
  \bibinfo{author}{\bibfnamefont{P.}~\bibnamefont{Ordej\'{o}n}},
  \bibnamefont{and}
  \bibinfo{author}{\bibfnamefont{D.}~\bibnamefont{Sanchez-Portal}},
  \bibinfo{journal}{J. Phys.: Condens. Matter} \textbf{\bibinfo{volume}{14}},
  \bibinfo{pages}{2745} (\bibinfo{year}{2002}).

\bibitem[{\citenamefont{Sankey and Niklewski}(1989)}]{Orbitals}
\bibinfo{author}{\bibfnamefont{O.~F.} \bibnamefont{Sankey}} \bibnamefont{and}
  \bibinfo{author}{\bibfnamefont{D.~J.} \bibnamefont{Niklewski}},
  \bibinfo{journal}{Phys. Rev. B} \textbf{\bibinfo{volume}{40}},
  \bibinfo{pages}{3979} (\bibinfo{year}{1989}).

\bibitem[{\citenamefont{Perdew and Wang}(1992)}]{GGA}
\bibinfo{author}{\bibfnamefont{J.~P.} \bibnamefont{Perdew}} \bibnamefont{and}
  \bibinfo{author}{\bibfnamefont{Y.}~\bibnamefont{Wang}},
  \bibinfo{journal}{Phys. Rev. B} \textbf{\bibinfo{volume}{45}},
  \bibinfo{pages}{13244} (\bibinfo{year}{1992}).

\bibitem[{\citenamefont{Berland and Hyldgaard}(2014)}]{BH}
\bibinfo{author}{\bibfnamefont{K.}~\bibnamefont{Berland}} \bibnamefont{and}
  \bibinfo{author}{\bibfnamefont{P.}~\bibnamefont{Hyldgaard}},
  \bibinfo{journal}{Phys. Rev. B} \textbf{\bibinfo{volume}{89}},
  \bibinfo{pages}{035412} (\bibinfo{year}{2014}).

\bibitem[{\citenamefont{Rom\'{a}n-P\'{e}rez and Soler}(2009)}]{Roman}
\bibinfo{author}{\bibfnamefont{G.}~\bibnamefont{Rom\'{a}n-P\'{e}rez}}
  \bibnamefont{and} \bibinfo{author}{\bibfnamefont{J.~M.} \bibnamefont{Soler}},
  \bibinfo{journal}{Phys. Rev. Lett.} \textbf{\bibinfo{volume}{103}},
  \bibinfo{pages}{096102} (\bibinfo{year}{2009}).

\bibitem[{SM()}]{SM}
\bibinfo{note}{See Suplemental Material}.

\bibitem[{\citenamefont{Methfessel and Paxton}(1989)}]{smearing}
\bibinfo{author}{\bibfnamefont{M.}~\bibnamefont{Methfessel}} \bibnamefont{and}
  \bibinfo{author}{\bibfnamefont{A.~T.} \bibnamefont{Paxton}},
  \bibinfo{journal}{Phys. Rev. B} \textbf{\bibinfo{volume}{40}},
  \bibinfo{pages}{3616} (\bibinfo{year}{1989}).

\bibitem[{\citenamefont{Li et~al.}(2010)\citenamefont{Li, Luican, dos Santos,
  Neto, Reina, Kong, and Andrei}}]{vanHove}
\bibinfo{author}{\bibfnamefont{G.}~\bibnamefont{Li}},
  \bibinfo{author}{\bibfnamefont{A.}~\bibnamefont{Luican}},
  \bibinfo{author}{\bibfnamefont{J.~M.} \bibnamefont{Lopes dos Santos}},
  \bibinfo{author}{\bibfnamefont{A.~C.} \bibnamefont{Neto}},
  \bibinfo{author}{\bibfnamefont{A.}~\bibnamefont{Reina}},
  \bibinfo{author}{\bibfnamefont{J.}~\bibnamefont{Kong}}, \bibnamefont{and}
  \bibinfo{author}{\bibfnamefont{E.}~\bibnamefont{Andrei}},
  \bibinfo{journal}{Nat. Phys.} \textbf{\bibinfo{volume}{6}},
  \bibinfo{pages}{109} (\bibinfo{year}{2010}).

\bibitem[{\citenamefont{Brihuega et~al.}(2012)\citenamefont{Brihuega, P.Mallet,
  Gonz\'{a}lez-Herrero, de~Laissardi\`{e}re, Ugeda, Magaud,
  G\'{o}mez-Rodr\'{i}guez, Yndurain, and Veuillen}}]{Nosotros-bilayer}
\bibinfo{author}{\bibfnamefont{I.}~\bibnamefont{Brihuega}},
  \bibinfo{author}{\bibnamefont{P.Mallet}},
  \bibinfo{author}{\bibfnamefont{H.}~\bibnamefont{Gonz\'{a}lez-Herrero}},
  \bibinfo{author}{\bibfnamefont{G.} \bibnamefont{Trambly~de~Laissardi\`{e}re}},
  \bibinfo{author}{\bibfnamefont{M.~M.} \bibnamefont{Ugeda}},
  \bibinfo{author}{\bibfnamefont{L.}~\bibnamefont{Magaud}},
  \bibinfo{author}{\bibfnamefont{J.}~\bibnamefont{G\'{o}mez-Rodr\'{i}guez}},
  \bibinfo{author}{\bibfnamefont{F.}~\bibnamefont{Yndurain}}, \bibnamefont{and}
  \bibinfo{author}{\bibfnamefont{J.-Y.} \bibnamefont{Veuillen}},
  \bibinfo{journal}{Phys. Rev. Lett.} \textbf{\bibinfo{volume}{109}},
  \bibinfo{pages}{196802} (\bibinfo{year}{2012}).

\bibitem[{\citenamefont{de~Laissardi\`{e}re
  et~al.}(2012)\citenamefont{de~Laissardi\`{e}re, Mayou, and
  Magaud}}]{Laissardiere-PRB}
\bibinfo{author}{\bibfnamefont{G.} \bibnamefont{Trambly~de~Laissardi\`{e}re}},
  \bibinfo{author}{\bibfnamefont{D.}~\bibnamefont{Mayou}}, \bibnamefont{and}
  \bibinfo{author}{\bibfnamefont{L.}~\bibnamefont{Magaud}},
  \bibinfo{journal}{Phys. Rev. B} \textbf{\bibinfo{volume}{86}},
  \bibinfo{pages}{125413} (\bibinfo{year}{2012}).

\bibitem[{\citenamefont{de~Laissardi\`{e}re
  et~al.}(2010)\citenamefont{de~Laissardi\`{e}re, Mayou, and
  Magaud}}]{Laissardiere1}
\bibinfo{author}{\bibfnamefont{G.} \bibnamefont{Trambly~de~Laissardi\`{e}re}},
  \bibinfo{author}{\bibfnamefont{D.}~\bibnamefont{Mayou}}, \bibnamefont{and}
  \bibinfo{author}{\bibfnamefont{L.}~\bibnamefont{Magaud}},
  \bibinfo{journal}{Nano Letters} \textbf{\bibinfo{volume}{10}},
  \bibinfo{pages}{804} (\bibinfo{year}{2010}).

\bibitem[{\citenamefont{dos Santos et~al.}(2007)\citenamefont{dos Santos,
  Peres, and Neto}}]{Continuum-1}
\bibinfo{author}{\bibfnamefont{J.~M. B.} \bibnamefont{Lopes dos Santos}},
  \bibinfo{author}{\bibfnamefont{N.~M.~R.} \bibnamefont{Peres}},
  \bibnamefont{and} \bibinfo{author}{\bibfnamefont{A.~H.}
  \bibnamefont{Castro Neto}}, \bibinfo{journal}{Phys. Rev. Lett.}
  \textbf{\bibinfo{volume}{99}}, \bibinfo{pages}{256802}
  (\bibinfo{year}{2007}).

\bibitem[{\citenamefont{dos Santos et~al.}(2012)\citenamefont{dos Santos,
  Peres, and Neto}}]{Continuum-2}
\bibinfo{author}{\bibfnamefont{J.~M. B.} \bibnamefont{Lopes~dos Santos}},
  \bibinfo{author}{\bibfnamefont{N.~M.~R.} \bibnamefont{Peres}},
  \bibnamefont{and} \bibinfo{author}{\bibfnamefont{A.~H.}
  \bibnamefont{Castro~Neto}}, \bibinfo{journal}{Phys. Rev. B}
  \textbf{\bibinfo{volume}{86}}, \bibinfo{pages}{155449}
  (\bibinfo{year}{2012}).

\bibitem[{\citenamefont{An and Pickett}(2001)}]{Pickett}
\bibinfo{author}{\bibfnamefont{J.~M.} \bibnamefont{An}} \bibnamefont{and}
  \bibinfo{author}{\bibfnamefont{W.~H.}~\bibnamefont{Pickett}},
  \bibinfo{journal}{Phys. Rev. Lett.} \textbf{\bibinfo{volume}{86}},
  \bibinfo{pages}{4366} (\bibinfo{year}{2001}).

\bibitem[{\citenamefont{Boeri et~al.}(2004)\citenamefont{Boeri, J.Kortus, and
  O.K.Andersen}}]{Andersen}
\bibinfo{author}{\bibfnamefont{L.}~\bibnamefont{Boeri}},
  \bibinfo{author}{\bibnamefont{J.Kortus}}, \bibnamefont{and}
  \bibinfo{author}{\bibnamefont{O.K.Andersen}}, \bibinfo{journal}{Phys. Rev.
  Lett.} \textbf{\bibinfo{volume}{93}}, \bibinfo{pages}{237002}
  (\bibinfo{year}{2004}).

\bibitem[{\citenamefont{Ramnani et~al.}(2017)\citenamefont{Ramnani, Neupane,
  Ge, Balandin, Lake, and a.~Mulchandani}}]{Phonons}
\bibinfo{author}{\bibfnamefont{R.}~\bibnamefont{Ramnani}},
  \bibinfo{author}{\bibfnamefont{M.}~\bibnamefont{Neupane}},
  \bibinfo{author}{\bibfnamefont{S.}~\bibnamefont{Ge}},
  \bibinfo{author}{\bibfnamefont{A.}~\bibnamefont{Balandin}},
  \bibinfo{author}{\bibfnamefont{R.~K.} \bibnamefont{Lake}}, \bibnamefont{and}
  \bibinfo{author}{\bibnamefont{A.}~\bibnamefont{Mulchandani}}, \bibinfo{journal}{Carbon}
  \textbf{\bibinfo{volume}{123}}, \bibinfo{pages}{302} (\bibinfo{year}{2017}).

\end{thebibliography}

\end{document}